 \documentclass[fleqn,11pt]{article}
 \usepackage{amsfonts,epsfig}
 \usepackage{times}

 \newcommand{\Ga}{\alpha}

 \newcommand{\Gd}{\delta}

 \newcommand{\Gl}{\lambda}
 \newcommand{\GL}{\Lambda}

 \newcommand{\CQ}{{\cal Q}}

 \newcommand{\ft}[2]{{\textstyle {\frac{#1}{#2}} }}

 \newcommand{\be}{\begin{equation}}
 \newcommand{\ee}{\end{equation}}
 \newcommand{\ben}{\begin{displaymath}}
 \newcommand{\een}{\end{displaymath}}
 \newcommand{\ba}{\begin{eqnarray}}
 \newcommand{\ea}{\end{eqnarray}}
 
 \newcommand{\non}{\nonumber\\}

 \newcommand{\mathon}{\mathversion{bold}}
 \newcommand{\mathoff}{\mathversion{normal}}

 \newcommand{\la}{\label}
 
 \newcommand{\Ref}[1]{(\ref{#1})}

 \newcommand{\g}{\mathfrak{g}}
 \newcommand{\h}{\mathfrak{h}}
 \newcommand{\n}{\mathfrak{n}}
 \newcommand{\3}{{\rm SO}(3)}
 \newcommand{\2}{{\rm SO}(1,2)}

 \newcommand{\equ}{\!=\!}

\begin{document}

 \thispagestyle{empty}

 \begin{flushright}
 AEI-2001-095\\
 SWAT-2001/312
 \end{flushright}
 \renewcommand{\thefootnote}{\fnsymbol{footnote}}

 \vspace*{0.1ex}
 \begin{center}
 \mathon
 {\bf\LARGE The Principal $\2$ Subalgebra of a Hyperbolic Kac Moody
 Algebra}
 \footnote{Work supported in part by the European Union under Contract
 No. HPRN-CT-2000-00131.}
 \mathoff
 \bigskip\bigskip\medskip

 {\bf {\large{H.~Nicolai}}}\medskip\\
 {\em Max-Planck-Institut f{\"u}r Gravitationsphysik,\\
   Albert-Einstein-Institut,\\
   M\"uhlenberg 1, D-14476 Golm, Germany}

 \smallskip {\small nicolai@aei-potsdam.mpg.de} \bigskip\smallskip
 \addtocounter{footnote}{-1}

 {\bf {\large{D.I.~Olive}}}\medskip\\
 {\em
 Department of Physics\\
 University of Wales Swansea\\
 Singleton Park\\
 Swansea SA2 8PP, UK}~
 \\

 \smallskip {\small d.i.olive@swansea.ac.uk\medskip}
 \end{center}
 \renewcommand{\thefootnote}{\arabic{footnote}}
 \setcounter{footnote}{0}
 \bigskip
 \bigskip

 \begin{abstract}
 The analog of the principal $\3$ subalgebra of a finite dimensional
 simple Lie algebra can be defined for any hyperbolic Kac Moody
 algebra $\g (A)$ associated with a symmetrizable Cartan matrix $A$,
 and coincides with the non-compact algebra $\2$. We exhibit
 the decomposition of $\g (A)$ into representations of $\2$; with
 the exception of the adjoint $\2$ algebra itself, all of these
 representations are unitary. We compute the Casimir eigenvalues;
 the associated ``exponents'' are complex and non-integer.
 \end{abstract}

 \renewcommand{\thefootnote}{\arabic{footnote}}
 \vfill

 \bigskip

 \leftline{{July 2001}}

 \setcounter{footnote}{0}
 \newpage

 \paragraph{Introduction.}
 Over the past years it has become clear that infinite dimensional
 Lie algebras \cite{Kac,MP} play an increasingly important role in
 modern theoretical physics, and string theory in particular. The
 close links between string vertex operators and Kac Moody algebras
 are well known \cite{BH,FK,GO}. In particular, there has been
 mounting evidence that Kac Moody algebras of indefinite type and
 generalized Kac Moody (Borcherds) algebras might appear in the
 guise of duality symmetries in string and M theory.

 In the context of gravity and supergravity, indications of a more
 concrete realization of indefinite, and especially hyperbolic, Kac Moody
 algebras have emerged very recently from results concerning the generic
 behavior of solutions of Einstein's field equations near a
 spacelike cosmological singularity (see \cite{DH1,DHJN} and
 references therein). More specifically, the diagonal gravitational
 metric degrees of freedom can be identified with the Cartan subalgebra
 of some underlying indefinite Kac Moody algebra, such that the resulting
 dynamics is elegantly described as a massless, relativistic,
 perfectly elastic billiard moving linearly within the associated
 fundamental Weyl chamber and bouncing off its walls. The presence
 of chaotic oscillations in this motion then becomes correlated to
 the hyperbolicity of this algebra.

 Possibly not unrelated is the fact that Kac Moody algebras
 have also appeared in Toda field theories (see
 \cite{OT,BCDS} and references therein). Hyperbolic Toda theories
 were considered in \cite{GIM}. Like the Toda theories based on
 finite dimensional Lie algebras, and in contradistinction to the
 affine Toda theories, they are conformally invariant due to
 the existence of a Weyl vector for the finite and hyperbolic cases.
 As also shown in \cite{GIM}, a special feature of the strictly
 hyperbolic Toda theories is the absence of higher order (spin 2 and
 above) local conserved charges, indicating the non-integrability of
 these models.

 As is well known, \cite{Kac,MP} every Kac Moody algebra is defined
 in terms of a Cartan matrix $A_{ij}$ ($1\leq i,j \leq r$) satisfying
 the properties listed on page 1 of \cite{Kac}, and a set of generating
 elements $\{e_i,f_i,h_i\}$ subject to a set of generating relations
 (Chevalley Serre presentation). The algebra itself can
 always be represented in the form
 \be
 \g (A) = \n_- \oplus \h \oplus \n_+
 \la{gA}
 \ee
 where $\h$ is the Cartan subalgebra (whose dimension equals the
 rank $r$), and where $\n_-$ and $\n_+$ are the triangular subalgebras
 consisting of the independent multiple commutators of the $e_i$'s
 and $f_i$'s, respectively. The most interesting (and least known)
 algebras are the ones associated with indefinite and non-degenerate
 Cartan matrices. Among them, {\it hyperbolic} Kac Moody algebras are
 distinguished by the extra requirement that the deletion of any node
 from the Dynkin diagram leaves a subalgebra which is either affine
 or finite. Unfortunately, beyond the usual ``Dynkinology'', they
 remain shrouded in mystery. Root multiplicities are known in closed
 form only for levels $|\ell |\leq 2$ \cite{FF,KMW}, and in implicit
 form also for $\ell =\pm 3$ \cite{BB}. While the Lie algebra
 elements at levels 0 and $\pm 1$ are completely under control,
 explicit representations of the root space elements beyond low
 levels have been worked out only for a few examples and exhibit
 an exceedingly complicated structure \cite{GN,GNW}. Especially in
 view of their conjectured applications, understanding the structure
 of hyperbolic Kac Moody algebras remains a major challenge.

 In this paper we take a step in this direction by generalizing
 a tool that has proved to be of great use in the study of finite
 dimensional simple Lie groups, namely the concept of the principal
 $\3$ subgroup \cite{Kostant} (its Lie algebra is distinguished
 amongst all those $\3$ subalgebras in the complete classification
 of \cite{Dynkin} by being maximal\footnote{For an early application
 of this classification in physics, see \cite{FS}; embeddings of the
 special superalgebra $OSp(1|2)$ into larger superalgebras were 
 studied in \cite{LSS}.}). The generalization of this concept to
 hyperbolic Kac Moody algebras rests on the fact that, among the set of
 all Cartan matrices, the finite and hyperbolic ones are singled out
 by the property that the entries of the associated inverse Cartan
 matrix have a definite sign, {\it i.e.} for all $i,j \in \{1,...,r\}$
 \ba
  A^{-1}_{ij} &\geq& 0 \quad \mbox{for finite Cartan matrices} \non
  A^{-1}_{ij} &\leq& 0 \quad \mbox{for hyperbolic Cartan matrices}
 \la{Cartan}
 \ea
 For definiteness and simplicity of notation, only symmetric Cartan
 matrices, $A$, are considered, and the real roots are assumed to
 have length $\sqrt2$; however, the extension of our results to the
 non-simply laced algebras based on symmetrizable Cartan matrices is
 straightforward. The second inequality in \Ref{Cartan} above follows
 from $A^{-1}_{ij} = \GL_i\!\cdot\!\GL_j$, and the fact that the
 fundamental
 weights $\GL_i$ all lie in the forward lightcone for hyperbolic
 Kac Moody algebras.

 The above properties have no analogue for affine Cartan matrices (which by
 definition are not invertible), and in general will also fail for
 non-degenerate indefinite Cartan matrices, where the entries of $A^{-1}$
 can assume both signs. We believe that we have thus found another
 indication of the privileged status enjoyed by the hyperbolic 
 algebras among the indefinite Kac Moody algebras.

 The main new insight of the present work is that, due to the relative
 switch in sign between finite and hyperbolic Cartan matrices
 $A$ in \Ref{Cartan}, the compact $\3$ associated with a finite
 dimensional Lie group is replaced by the non-compact group $\2$ for
 hyperbolic Kac Moody algebras. With the exception of the adjoint
 representation, the finite dimensional representations of $\3$ are
 accordingly replaced by infinite dimensional ones. We will exhibit
 the basic features arising when the algebra $\g (A)$ is decomposed
 into representations of $\2$. The fact that the ``exponents'' of
 $\g (A)$ now come out to be complex and irrational is presumably
 related to the fact that hyperbolic Kac Moody algebras do not admit
 any polynomial Casimir invariants other than the quadratic Casimir
 Kac operator \cite{Kac1}.

 \paragraph{The principal SO(3) subalgebra of a finite Kac Moody algebra.}

 This algebra exists for every Kac Moody algebra defined by a
 positive definite Cartan matrix $A_{ij}$ (it is a standard result
 that the positive definiteness of $A$ implies that $\g (A)$
 is finite dimensional \cite{Kac,MP}). It is constructed
 by means of the Weyl vector $\rho$, which is defined to obey
 $\rho\!\cdot\!\Ga_i = 1$ for all simple roots $\Ga_i$. An explicit
 formula is $\rho = \sum_j \GL_j$ where $\GL_j$ are the fundamental
 weights satisfying $\Ga_i\!\cdot\!\Lambda_j = \Gd_{ij}$. The diagonal
 generator of the principal $\3$ is defined by
 \be
 J_3 := \rho \!\cdot\! H \;\;\; \Longrightarrow \;\;\;
 [J_3, E_\Ga] = {\rm ht}(\Ga)\, E_\Ga
 \la{J3}
 \ee
 where ${\rm ht}(\Ga)$ denotes the height of the root $\Ga$ and
 $E_\Ga$ the generator(s) associated with the root $\Ga$. Then,
 since the number of simple roots equals the dimension of the
 Cartan subalgebra, $r$,
 there always exist linear combinations of the step operators for
 the simple roots
  $E_{\Ga_i}$ and $E_{-\Ga_i}$
 \be
 J^+ = \sum n_i E_{\Ga_i} \;\; , \quad
 J^- = \sum n_i E_{-\Ga_i}
 \ee
 such that
 \be
 [J_3, J^\pm] = \pm J^\pm \;\; , \quad [J^+,J^-] = J_3
 \la{SO3}
 \ee

 With respect to the principal $\3$ algebra, the Lie algebra $\g (A)$
 decomposes into $r$ irreducible representations of spin $s_j$
 \be
 \g (A) = \bigoplus_{j=1}^r \g^{(s_j)}
 \ee
 where $\g^{(s_j)}$ carries the $(2s_j+1)$-dimensional irreducible
 representation of $\3$, and the $r$ ``spins''
 $s_j$ are known as the exponents of $\g (A)$. In particular,
 $\g^{(0)}$ is empty, while the adjoint representation $\g^{(1)}$ is
 just the principal $\3$ subalgebra itself. Thus the smallest exponent
 is always $s_1=1$. The importance of the
 principal $\3$ is due to the fact that the exponents $s_j$
 contain essential information about the Lie group. For instance,
 the orders of the invariant Casimir operators are given by
 the numbers $s_j+1$; thus, the representation $s_1\equ 1$ is always
 associated with the quadratic Casimir invariant. Furthermore,
 the group (co)homology is specified by the Poincar\'e polynomial
 $\prod_j (1 - x^{2s_j +1})$ \cite {Kostant}.

 It is straightforward to re-express the $\3$ generators directly in
 terms of the Chevalley basis $e_i\equiv E_{\Ga_i}, f_i\equiv E_{-\Ga_i}$
 and $h_i \equiv \Ga_i\!\cdot\! H$. Using $\GL_i = \sum_j A^{-1}_{ij}
 \Ga_j$
 in \Ref{J3} we readily obtain
 \be
 J_3 = \sum_j p_j h_j \; , \quad J^+ = \sum_j n_j e_j \; , \quad
  J^- = \sum_j n_j f_j
 \la{Chevalley}
 \ee
 where
 \be
 p_i := \sum_j A^{-1}_{ij} > 0 \;\; , \qquad n_i:= \sqrt{p_i}
 \la{p1}
 \ee
 The strict positivity of $p_i$ for all $i$ follows from \Ref{Cartan}
 and the non-degeneracy of $A_{ij}$. The $\3$ algebra can now be
 directly verified from the standard Chevalley-Serre presentation.

 \paragraph{The principal SO(1,2) subalgebra of a hyperbolic
 Kac Moody algebra.}
 Because the Weyl vector exists also for certain infinite dimensional
 Kac Moody algebras, it is natural to extend the above considerations
 to Kac Moody algebras whose Cartan matrices $A$ are no longer positive
 definite. However, the mere existence of a Weyl vector by itself
 is not sufficient; rather, it is the fact that the entries of the
 inverse Cartan matrix are of the same sign which ensures that the
 construction can be carried through. For the hyperbolic case the
 expression for $J_3$ still takes the same form $\sum_{i,j}A^{-1}_{ij}h_j$ 
 as before. But, taking account of the relative sign switch in \Ref{Cartan} 
 and insisting that $p_j$ still denotes a positive quantity, we now have,
 instead of \Ref{Chevalley} and \Ref{p1},
 \be
 J_3 = - \sum_j p_j h_j
 \la{p2}
 \ee
 with
 \be
 p_i := -\sum_j A^{-1}_{ij} > 0 \;\; , \qquad n_i:= \sqrt{p_i}
 \ee
 As a consequence of the extra minus signs in these definitions,
 \Ref{SO3} is replaced by
 \be
 [J_3, J^\pm] = \pm J^\pm \;\; , \quad [J^+,J^-] = - J_3
 \ee
 Because the hermiticity properties of the Chevalley generators are
 the same as before, we see that the compact $\3$ has been replaced
 by a non-compact $\2$. Evidently, an analogous definition cannot
 work for affine Cartan matrices, whose inverse does not exist;
 likewise, it fails for Kac Moody algebras where the
 signs of the $p_i$'s alternate. The consistency of the above
 definition may be traced back in part to the fact that the Weyl
 vector is timelike ($\rho^2 <0$) and an element of the forward
 lightcone in root space for hyperbolic Cartan matrices (this property
 actually holds for all indefinite algebras of rank $\leq 25$,
 provided they are obtained by the procedure of overextension \cite{GO}).

 As before it is possible to decompose the algebra $\g (A)$ into
 irreducible representations of the principal subalgebra. However,
 in accordance with the non-compactness of $\2$, all the irreducible
 representations occurring will now be infinite-dimensional and unitary,
 with the  exception of  the adjoint representation consisting
 of the subalgebra $\2$ itself, which is neither.

 In this context, according to standard definitions, \cite{Kac},  
 unitary means that
 the representation space possesses a hermitian scalar product, denoted
 $(x,y)$ with the properties that (i) the actions of $e_i$ and $f_i$
 are mutually adjoint, while that of $h_i$ is selfadjoint, {\it i.e.}
 for all $x,y\in \g(A)$
 \be
 \big([e_i,x],y\big) = \big(x,[f_i,y]\big) \quad \mbox{and} \quad
 \big([h_i,x],y\big) = \big(x,[h_i,y]\big)
 \la{adjoint}
 \ee
 and (ii) the scalar product is positive definite.

 Here the representation space is the vector space of the algebra
 $\g(A)$, with $\g(A)$ acting on itself by adjoint action.
 Because the Cartan matrix $A$ is assumed symmetric the algebra possesses
 a standard invariant bilinear form $\langle.|.\rangle$, generalising
 the Cartan Killing form.
 A natural candidate for the hermitian scalar product
 (extending that familiar in angular momentum theory) is given by 
 (\cite{Kac}, chapter 2),
 \be
 (x,y) := -\langle x|\theta(y) \rangle
 \la{scalarproduct}
 \ee
 where $\theta$ is the Chevalley involution
 \be \theta (e_i) = - f_i \;\; , \quad
     \theta (f_i) = - e_i \;\; , \quad
     \theta (h_i) = - h_i
 \ee
 It is easy to check that with this definition $e_i$ and $f_i$ are
 indeed mutually adjoint while $h_i$ is selfadjoint but the question
 of positive definiteness is more subtle. There is a rather general
 theorem,
 \cite{KP1}, certainly applicable to hyperbolic algebras with symmetric
 Cartan matrix, that states that $\g(A)$ as a vector space decomposes
 into orthogonal subspaces consisting of the Cartan subalgebra and
 subspaces associated with each root. All subspaces are positive definite
 with respect to \Ref{scalarproduct} except for the Cartan subalgebra for
 which the scalar product reduces to the indefinite one
 already met in talking of scalar products between roots and weights.

 Having verified the desired adjointness properties of the Chevalley
 generators, it follows immediately from \Ref{Chevalley} that $J_3$ is
 selfadoint while $J^+$ and $J^-$ are mutually adjoint. Furthermore,
 the norms of these elements are easily calculated to be
 $(J_3,J_3)=\rho^2= - \sum_j p_j < 0$ and
 $(J^-,J^-) = (J^+ , J^+) = \sum p_j > 0$. Thus the  adjoint
 representation of $SO(1,2)$ is indeed not unitary. The reason is that
 the Weyl vector $\rho$ is inside the forward light cone. Since all
 vectors orthogonal to it are space-like the hermitian scalar product
 restricted to this  subspace within the Cartan subalgebra orthogonal
 to the Weyl vector is positive definite. Because the decomposition of
 $\g(A)$ into irreducible representations is into orthogonal subspaces
 this is the reason that all the components except the three dimensional
 one are unitary.

 \paragraph{Irreducible representations of SO(1,2).}
 Next we examine which sorts of unitary representations of $SO(1,2)$ occur.

 Because of the adjoint action the spectrum of $J_3$ is integral, that
 is $\exp (2\pi i J_3) $ equals unity, the representations arising must
 be what is sometimes called single valued, as well as unitary.
 As usual, the irreducible representations of $\2$ are labeled
 (in part) by the eigenvalues of the $\2$ Casimir operator
 \ba
 \CQ &=& J_3 J_3 - J^+ J^- - J^- J^+ \non
     &=& J_3 (J_3-1) - 2J^+ J^- =  J_3 (J_3+1) - 2 J^- J^+
 \la{Casimir1}
 \ea
 When evaluating this Casimir on a given element $x\in \g (A)$
 we will always understand the adjoint action
 \be
 {\rm ad}_\CQ \, (x) := [J_3,[J_3,x]] - [J^+,[J^-,x]] - [J^-,[J^+,x]]
 \la{adjointaction}
 \ee

 Besides the non-unitary finite dimensional representations such
 as the three-dimensional one,
 $\2$ possesses two different kinds of unitary infinite dimensional
 representations \cite{BF}. The so-called discrete series
 representations with Casimir eigenvalue $\CQ = s(s-1) > 0$ are
 characterized by the existence of a lowest (highest) weight state obeying
 $J^- |s,s\rangle = 0$ (or  $J^+ |-s,-s\rangle = 0$); the states
 of the representation are then given by $|s,m\rangle$ (or
 $|-s,-m\rangle$)
 for $m=s,s+1,\dots$. Because we are interested only in the adjoint
 action \Ref{adjointaction} we encounter only single-valued
 representations, namely ones obeying $\exp (2\pi i J_3) = 1$.
 Hence for the discrete series occurring in the decomposition
 of $\g (A)$, $s$ is always a (positive) integer --- in fact,
 we will see below that actually $s\geq 2$. The continuous
 representations split into principal and supplementary series
 representations, with the respective  Casimir eigenvalues obeying
 \ba
 \mbox{principal series:} \qquad & -\infty < \CQ < -\ft14 &  \non
 \mbox{supplementary series:} \qquad & -\ft14 < \CQ < 0  &
 \la{CQ}
 \ea
 These inequalities are easily deduced from \Ref{Casimir1} and the
 positivity requirements $\big( x, [J^\pm, [J^\mp,x]]\big) >0$ for
 $x\neq 0$, implying $\CQ < m(m \pm 1)$ for all integer-spaced
 $m \in E_0 + {\bf Z}$. For the supplementary series $m$ is never 
 an integer, and therefore the latter representations will not occur 
 in our analysis since $\exp (2\pi i J_3)$ cannot equal unity for them.

 To analyze the representation content of $\g (A)$, let us first
 consider those on spaces intersecting the Cartan subalgebra $\h$.
 Apart from $J_3$, there are $(r-1)$ linearly independent combinations
 belonging to principal series representations. For any linear combination
 $\sum_j c_j h_j$ we have
 \be
 {\rm ad}_\CQ \, \Big(\sum_j c_j h_j \Big) = -2 [J^- , [J^+ , \sum_j c_j
 h_j ]]
     = - 2 \sum_{i,j} c_i A_{ij} p_j h_j
 \la{Casimir}
 \ee
 Setting $c_j = p_j$ and using $\sum_j A_{ij} p_j = -1$ for all $i$,
 we obtain
 \be
 {\rm ad}_\CQ \, \big(J_3 \big) = -2 \sum_{i,j} p_i A_{ij} p_j h_j = +2
 J_3
 \ee
 as expected for the adjoint representation of $\2$. We have already
 mentioned that the latter is the only finite dimensional representation arising.
 The coefficients of the $(r-1)$ orthogonal linear combinations satisfy
 \be
 \Big(\sum_i c_i' h_i , \sum_j p_j h_j\Big) =
 \sum_{i,j} c_i' A_{ij} p_j = 0 \;\; \Longrightarrow
 \quad \sum_j c_j' = 0
 \ee
 It is not difficult to see that these orthogonal combinations are
 of positive norm because the Weyl vector is timelike, and therefore
 any vector $\sum_j c_j' \Ga_j$ orthogonal to it must be spacelike.
 We can now generate the full representations by multiply commuting
 $\sum_j c_j' h_j$ with $J^+$ and $J^-$, where the $(r-1)$ mutually
 orthogonal linear combinations  $\sum_j c_j' h_j$ are determined
 by diagonalizing the $\2$ Casimir operator \Ref{Casimir}.
 Since none of these commutators vanishes, these representations
 extend simultaneously into $\n_-$ and $\n_+$. For instance,
 commuting once with $J^+$ we obtain
 \be
 x:= [J^+, \sum_j c_j' h_j] = - \sum_{i,j} c_i' A_{ij} n_j e_j
 \ee
 and one easily checks that $(x,x)>0$. The positivity of the
 remaining states in the representation then follows from the theorem
 mentioned above which is indeed proven by induction on the height
 of the roots \cite{KP1}. Because the (integer)
 eigenvalues of $J_3$ are bounded neither from below nor from above,
 we conclude that the orthogonal complement of $J_3$ in $\h$ must
 belong to $(r-1)$ principal series representations.

By contrast, the discrete series representations are entirely contained 
in the triangular subalgebras $\n_+$ or $\n_-$. The lowest weight 
representations are built on states of the form 
\be 
v^{(s)}=\sum_{j_1\dots j_s} c_{j_1\dots j_s} 
           \, [e_{j_1}, \dots ,[e_{j_{s-1}},e_{j_s}]\dots]
\; , \qquad \big[J^- , v^{(s)}\big] = 0
\la{lw}
\ee
by repeated application of $J^+$. In an analogous fashion, the 
highest weight states are obtained by acting on the states 
\be 
v^{(-s)}=\sum_{j_1\dots j_s}  c_{j_1\dots j_s} 
           \, [f_{j_1}, \dots ,[f_{j_{s-1}},f_{j_s}]\dots]
\; , \qquad \big[J^+ , v^{(-s)}\big] = 0
\la{hw}
\ee
with $J^-$. From \Ref{J3} it is immediately obvious that the
lowest weight states indeed have spin $s$ \footnote{Alternatively,
we may use the formula
$$
\big[h_i\, ,\, [e_{j_1}, \dots ,[e_{j_{s-1}},e_{j_s}]\dots] \big]
= \left( \sum_{k=1}^s A_{ij_k} \right)
 [e_{j_1}, \dots ,[e_{j_{s-1}},e_{j_s}]\dots]
$$
which is easily proved by induction.}. Because the space 
spanned by the generators $e_i$ is of dimension $r$, there are 
no new representations at that level (corresponding to spin $s=1$). 
Likewise, for $s=2$, the number of independent Lie algebra 
elements of type $[e_i,e_j]$, that is corresponding to roots of 
height two, equals the number of links in the Dynkin diagram, 
which is at most  $r$ for hyperbolic diagrams. Hence, at most one new 
representation starts with $s=2$, and that only if the diagram has 
a loop rather than a tree structure. Thus only the spins 
$s=2,3,4,\dots$ with corresponding Casimir eigenvalues $\CQ=s(s-1)$ 
occur in the discrete series representations, whose unitarity follows 
again by the general theorem. We emphasize that the discrete series 
representations have no analog in the finite dimensional case, where 
all representations appearing in the decomposition of the Lie algebra 
intersect the Cartan subalgebra non-trivially. 

 \paragraph{Casimir eigenvalues.}
 We now wish to calculate the Casimir eigenvalues of the principal
 series representations occurring in the decomposition of $\g (A)$
 for some concrete examples. These representations are the
 the analogues of the $r$ representations occurring in the
 decomposition of a finite dimensional Lie algebra. Of these,
 the finite dimensional adjoint representation with $\CQ = +2$
 is present in both the finite and the infinite dimensional case,
 and is unitary for $\3$, and non-unitary for $\2$. The remaining
 $(r-1)$ representations belonging to the principal series
 must satisfy the bound $\CQ < -\ft14$. Setting
 \be
 \CQ = s_j(s_j-1) \quad  \mbox{for $j=2,\dots,r$}
 \ee
 we have
 \be
 s_j= \ft12 + i\Gl_j
 \la{exponents}
 \ee
 The resulting $(r-1)$ values of $s_j$ can be viewed as the analogs
 of the exponents in the finite dimensional case, but they are
 now complex and non-integer.

 From \Ref{Casimir}, we infer that the Casimir eigenvalues of
 the adjoint and principal series representations are identical
 with the eigenvalues of the non-symmetric real matrix $-2 A_{ij} p_j$
 (no summation on $j$). This matrix was actually introduced already
 in \cite{GIM}, albeit for the (slightly different) purpose of
 determining the location of ``resonances'' in the associated
 strictly hyperbolic Toda models. Besides the infinitely many
 strictly hyperbolic Kac Moody algebras of rank two\footnote{With
 Cartan matrices (for $mn>4$)
 $$ 
 A_{ij} = \pmatrix{ 2  &-m \cr
                    -n & 2 \cr}
 $$}
 there are altogether eleven such algebras of rank 
 three and four, see e.g. \cite{S}.
 The relevant eigenvalues are listed in Table 3 of \cite{GIM},
 and with the benefit of hindsight are now easily recognized
 to be just the $\2$ Casimir eigenvalues for these algebras.

 The simplest hyperbolic algebra obtained by over-extension (hence
 containing an affine subalgebra) is $AE_3$, which was first
 studied in \cite{FF}. In this case, $\{p_j\} = (\,\ft92\,|\,5\,|\,2\,)$,
 and therefore
 \ben
 A_{ij} p_j = \pmatrix{ 9 &-10 &  0 \cr
                       -9 & 10 & -2 \cr
                        0 & -5  & 4 \cr}
 \een
 The eigenvalues of $-2 A_{ij} p_j$ are given by
 \be
 \CQ = 2 \, ,\, -2\left(12 \pm \sqrt{54}\right)
 \ee
 In a similar manner one determines the Casimir eigenvalues of the
 hyperbolic algebras $AE_n$ for $n>3$.

 For the maximally extended hyperbolic algebra $E_{10}$, we have
 \be
 \{p_j\}=
 (\,30\,|\,61\,|\,93\,|\,126\,|\,160\,|\,195\,|\,231\,|\,153\,|\,76\,|\,115\,),
 \ee
 and the matrix $A_{ij} p_j$ is
 \medskip
 \ben
 \pmatrix{60 & -61 & 0 & 0 & 0& 0& 0& 0& 0& 0 \cr
         -30 & 122 & -93 & 0& 0& 0& 0& 0& 0& 0 \cr
          0 & -61 & 186 & - 126 & 0& 0& 0& 0& 0& 0 \cr
          0 & 0 & -93 & 252 & -160& 0& 0& 0& 0& 0 \cr
          0 & 0 & 0 & - 126 & 320& -195 & 0& 0& 0& 0 \cr
          0 & 0 & 0 & 0 & -160& 390& -231& 0& 0& 0 \cr
          0 & 0 & 0 & 0 & 0& -195& 462& -153& 0& -115 \cr
          0 & 0 & 0 & 0 & 0& 0& -231& 306& -76 & 0 \cr
          0 & 0 & 0 & 0 & 0& 0& 0& -153 & 152 & 0 \cr
          0 & 0 & 0 & 0 & 0& 0& -231& 0 & 0 & 230 \cr  }
 \een
 \medskip
 The eigenvalues of  $-2 A_{ij} p_j$ can be determined
 numerically. Besides the expected eigenvalue $\CQ=2$,
 we find the nine values
 \ba
 \CQ &=& - 45.86857088 \non
     && - 124.74542658 \non
     && -221.4130766  \non
     && -290.5176114 \non
     && -438.1539904  \non
     && -594.5608986  \non
     && -714.1355888  \non
     && - 1025.0975582 \non
     && - 1507.5072788
 \la{eigenvalues}
 \ea
 The determination of the ``exponents'' \Ref{exponents} is now an
 elementary exercise. The complexity of these numbers is related
 to the non-existence of polynomial invariants in the enveloping
 algebra of $\g (A)$ other than the quadratic Casimir-Kac element
 \cite{Kac,MP}. However, as shown in \cite{Kac1}, there do exist
 transcendental invariant functions on the Cartan subalgebra. The
 precise link between them and the exponents exhibited above
 remains to be elucidated, however.

 \paragraph{Outlook.}

 Whereas the principal series representations are uniquely determined
 by diagonalizing the $\2$ Casimir
 operator, it is less easy to describe the spectrum of discrete
 series representations. Certainly the number of highest
 (or lowest) weight states will increase
 exponentially with the height and (negative) length of the roots, leaving
 an equally growing arbitrariness in the number of ways they can be
 combined into linearly independent and mutually orthogonal elements.

 On the other hand, we expect the states belonging to the principal series
 representations to play a distinguished role, and to provide a new
 way of ``probing'' hyperbolic Kac Moody algebras. Usually, the hyperbolic
algebras which arise as over-extensions of affine algebras, are
 decomposed w.r.t. to the level \cite{FF,KMW} (defined as the eigenvalue
 of the central charge operator of the underlying affine algebra), viz.
 \be
 \g (A) = \bigoplus_{\ell\in{\bf Z}} \g^{[\ell]}(A)
 \ee
 Because the generators $J^\pm$ always have a contribution from
 the over-extended root, we see that their action does not preserve
 the level. For this reason, in any of the $(r-1)$ principal series
 representations, there will be states mixing an arbitrary (but given)
 number of levels. Therefore the decompositions w.r.t. to level and
 w.r.t. to $\2$ are extremely oblique relative to one another.

\medskip 

\noindent
{\bf Acknowledgment.} D.I.~Olive would like to thank the 
Albert Einstein Institute for hospitality during his stay there.
We are also grateful to T.~Fischbacher for help with the numerical
computations.


\begin{thebibliography}{10}

 \bibitem{Kac} V.G. Kac, {\it Infinite Dimensional Lie Algebras},
 third edition (Cambridge University Press, 1990).
 %
 \bibitem{MP} R.V.~Moody and A.~Pianzola, {\it Lie Algebras with
 Triangular Decomposition} (John Wiley and Sons, New York, 1995)
 %
 \bibitem{BH} K.~Bardak\c{c}i and M.B.~Halpern,
 Phys. Rev. {\bf D3} (1971) 2493
 %
 \bibitem{FK} I.B.~Frenkel and V.G.~Kac, Invent. Math. {\bf 62} (1980) 23
 %
 \bibitem{GO} P.~Goddard and D.I.~Olive, in {\it Vertex Operators in
 Mathematics and Physics}, eds. J.~Lepowsky et al., MSRI Publication Nr.3,
 Springer Verlag, Heidelberg, 1985) 51--96
 %
 \bibitem{DH1} T.~Damour and M.~Henneaux,
 Phys. Rev. Lett. {\bf 86} (2001) 4749 [hep-th/0012172]
 %
 %
 \bibitem{DHJN} T.~Damour, M.~Henneaux, B.~Julia and H.~Nicolai,
 Phys. Lett. {\bf B509} (2001) 323, [hep-th/0103094]
 %
 \bibitem{OT} D.I.~Olive and N.~Turok, Nucl. Phys. {\bf B215}[FS7] (1983)
 470; Nucl. Phys. {\bf B257}[FS14] (1985) 277
 %
 \bibitem{BCDS}
 H.W.~Braden, E.~Corrigan, P.E.~Dorey and R.~Sasaki,
 Nucl. Phys. {\bf B338} (1990) 689; {\bf B356} (1991) 469
 %
 \bibitem{GIM} R.W.~Gebert, T.~Inami and S.~Mizoguchi,
 Int. J. Mod. Phys. {\bf A11} (1996) 5479 [hep-th/9503176]
 %
 \bibitem{FF} A.~Feingold and I.~Frenkel,
 Math. Ann. {\bf 263}  (1983) 87.
 %
 \bibitem{KMW} V.G.~Kac, R.V.~Moody and M.~Wakimoto, in {\it Differential
 Geometrical methods in Theoretical Physics}, Proc. NATO Advanced Research
 Workshop, 16th Int. Conf., Como, eds. K.~Bleuler and M.~Werner
 (Kluwer, Holland), 109--128
 %
 \bibitem{BB} M.~Bauer and D.~Bernard, Lett. Math. Phys. {\bf 42} (1997) 153,
     [hep-th/9612210]
 %
 \bibitem{GN} R.W.~Gebert and H.~Nicolai,
 Commun. Math. Phys. {\bf 172} (1995) 571
 %
 \bibitem{GNW} R.W.~Gebert, H.~Nicolai and P.C.~West,
 Int. Journ. Mod. Phys. {\bf A11} (1996) 429--514
 %
 \bibitem{Kac1} V.G.~Kac, Proc. Natl. Acad. Sci. USA {\bf 81} (1984) 645
 %
 \bibitem{Kostant} B.~Kostant, Am. J. Math. {\bf 81} (1959) 973
 %
 \bibitem{Dynkin} E.B.~Dynkin, Translat. Am. Math. Soc.Ser. 2 {\bf 6} 
         (1957) 111
 %
 \bibitem{FS}
 M.~Flato and D.~Sternheimer, J. Math. Phys. {\bf 7} (1966) 1932
 %
 \bibitem{LSS} D.A.~Leites, M.V.~Saveliev and V.V.~Serganova, in:
 Group theoretical methods in physics, Vol. I (Yurmala, 1985) 255,
 VNU Sci. Press, Utrecht, 1986.
 %
 \bibitem{BF} A.O.~Barut and C.~Fronsdal, Proc. Royal Soc. {\bf A287}
 (1965) 532
 %
 \bibitem{KP1} V.G.~Kac and D.H.~Peterson, Invent. Math. {\bf 76} (1984) 1
 %
 \bibitem{S} C.~Sa\c{c}lio\u{g}lu, J. Phys. {\bf A22} (1989) 3753
 %
 \end{thebibliography}

 \providecommand{\href}[2]{#2}\begingroup\raggedright\endgroup

 \end{document}